# X-ray powder diffraction of high-absorption materials at the XRD1 beamline off the best conditions: Application to (Gd,Nd)$_5$Si$_4$ compounds

A. M. G. Carvalho[1], R. S. Nunes[1,2], A. A. Coelho[3]

[1]*Laboratório Nacional de Luz Síncrotron, CNPEM, 13083-970, Campinas, SP, Brazil.*
[2]*Unidade Acadêmica de Física, UFCG, 58429-900, Campina Grande, PB, Brazil.*
[3]*Instituto de Física Gleb Wataghin, UNICAMP, 13083-859, Campinas, SP, Brazil.*

**Abstract**

Representative compounds of the new family of magnetic materials Gd$_{5-x}$Nd$_x$Si$_4$ were analyzed by X-ray diffraction at the XRD1 beamline at LNLS. To reduce X-ray absorption, thin layers of the powder samples were mounted outside the capillaries and measured in Debye-Scherrer geometry as usual. The X-ray diffraction analyses and the magnetometry results indicate that the behavior of the magnetic transition temperature as a function of Nd content may be directly related to the average of the four smallest interatomic distances between different rare earth sites of the majority phase of each compound. The quality and consistency of the results show that the XRD1 beamline is able to perform satisfactory X-ray diffraction experiments on high-absorption materials even off the best conditions.

# I. Introduction

The Laboratório Nacional de Luz Síncrotron (LNLS), also known as the Brazilian Synchrotron Light Laboratory, was the first synchrotron facility in the southern hemisphere and is the only one in Latin America. It is a second-generation synchrotron source, which operates with the energy of 1.37 GeV. One of the beamlines dedicated to X-ray diffraction is XRD1, which is installed on D12B bending magnet (1.67 T) and operates in the 5.5–14 keV (2.25–0.89 Å) range. Recently, the XRD1 beamline was upgraded (Canova et al., 2014) and its commissioning was entirely performed.

XRD1 beamline has two end-stations since 2010. The second end-station consists of a facility built around an advanced thermo-mechanical simulator, the Gleeble®Synchrotron system, which allows the material of interest to be submitted to a wide range of thermo-mechanical conditions (XTMS Experimental Station, 2015). The first end-station was built in 1997. With the upgrade process, the old diffractometer was exchanged by the 3-circle Heavy Duty diffractometer from Newport® (Canova et al., 2014; Tang et al., 2009). Furthermore, the installation of the MYTHEN 24K system, from Dectris®, and a robotic arm sample changer allowed relatively fast X-ray powder diffraction measurements. The measurement times vary typically between 30 and 400 seconds, depending on the sample and the investigation aims.

Initially, the XRD1 first end-station was optimized for X-ray diffraction measurements of low-absorption materials in powder form, using glass capillaries in Debye-Scherrer geometry. The typical diameters of the capillaries vary between 0.3 and 1.0 mm. However, even 0.3-mm-diameter capillaries are too large for high-absorption materials, because of X-ray absorption and the limitation in energy and photon flux on XRD1 beamline (Canova et al., 2014). To overcome this issue, we have been using the simple methodology of fixing the samples powder outside the capillaries (see, for instance, Pecharsky and Zavalij, 2005).

In the present work, we show the ability of XRD1 beamline to perform satisfactory X-ray powder diffraction off the best condition of the beamline and samples preparation. Then, we have chosen to analyze high-absorption materials during the beamline commissioning. The powder of the two-phase $GdNd_4Si_4$ compound was studied by X-ray diffraction inside the 0.3-mm-diameter capillary and outside capillaries with different diameters. Furthermore, representative compounds of the new family of magnetic materials $Gd_{5-x}Nd_xSi_4$ were analyzed by X-ray diffraction and magnetometry. We have observed a correlation between the magnetic transition temperatures and the average of the four smallest interatomic distances between different rare earth sites in the majority phases.

## II. Experimental procedures

Polycrystalline samples of compounds with nominal composition corresponding to $Gd_{5-x}Nd_xSi_4$ ($x$ = 1, 2, 3, 4 and 5) were prepared by arc-melting the elements in high purity argon atmosphere on a water-cooled copper hearth. The purity of the starting elements was 99.9 wt.% for Gd and Nd and 99.99+ wt.% for Si. The samples were annealed at 1050 °C for 10 days.

For X-ray diffraction measurements, each annealed sample was manually ground in an agate mortar. The obtained powder was not sieved purposely. A 0.3-mm-diameter capillary was filled, as usual, with the powder of $GdNd_4Si_4$ compound. The powder was also fixed outside borosilicate glass capillaries with outer diameters of 0.3 mm, 0.5 mm and 0.7 mm, using a fine layer of a commercial Vaseline. The capillaries have wall thickness of 0.01 mm and were supplied by Capillary Tube Supplies Ltd. We name four experimental settings: *setting I* - powder inside 0.3-mm-diameter capillary; *setting II* - powder outside 0.3-mm-diameter capillary; *setting III* - powder outside 0.5-mm-diameter capillary; *setting IV* - powder outside 0.7-mm-diameter capillary. The powder excess is removed using a vibrating device developed in-house. The powders of the other compounds of the series were fixed only outside the 0.7-mm-diameter capillaries. The capillaries are mounted in ferromagnetic stainless steel holders. The capillary holder is fit together with the magnetic tip attached to the 3-circle diffractometer. This magnetic tip is able to rotate, providing the desirable spinning (~ 300 r.p.m.) of capillaries during the measurements. The X-ray beam size at the sample position is ~ 2.0 mm (horizontal) and ~ 0.7 mm (vertical). During the measurements, the capillaries rely on the horizontal plane, along the direction perpendicular to the beam, at 760 mm from the detectors.

The X-ray diffraction patterns were obtained at room temperature using the MYTHEN 24K system, from Dectris®. This kind of position-sensitive detectors have been used for a few years in other synchrotron facilities (see, for instance, Bergamaschi et al., 2010; Thompson et al., 2011). Since there is a gap between each detector, a whole diffraction pattern is, in fact, a merging of two diffraction patterns obtained in two different positions 0.5° apart from each other. The step size is fixed at 0.004°. The NIST SRM640d standard Si powder was used to determine the X-ray wavelength with precision. The wavelength used was 1.0328(1) Å. Additional information of the beamline: ULE (ultra-low expansion) Rh-coated glass mirror, which is used to focus/collimate the white beam vertically, as well as filter the high-energy photons; double-crystal Si (111) monochromator; energy resolution $\Delta E/E = 3 \times 10^{-4}$ at 8 keV (1.55 Å).

The magnetic transition temperatures were obtained from the magnetization measured as a function of the temperature using the commercial MPMS magnetometer, from Quantum Design®. The magnetic measurements were performed on the annealed pieces of the studied compounds. The values of the transition temperatures were determined from the derivative minimum in the magnetization vs. temperature curves.

### III. Results and discussion

The X-ray diffraction patterns were analyzed using *TOPAS* software (Bruker: TOPAS software, 2015). Structural analysis using the Rietveld refinement method (Rietveld, 1969) achieved satisfactory values, refining zero error and 6 terms of Chebychev background polynomial as general independent parameters; for each crystal structure: one scale factor, lattice parameters (3 for orthorhombic and 2 for tetragonal), TCHZ peak type (the first 4 terms of peak shape) and atomic displacement parameters (depending on atomic sites). The annealed samples are not single-phase. Only for $Gd_4NdSi_4$ compound, the reflections of the secondary phase were not considered in the refinement because their very low intensities. Good fits were achieved from Rietveld refinement analysis for all settings (see Fig. 1b, for fit of *setting II*). For *setting II*, we have obtained $R_{wp} = 9.19\%$, $gof = 3.221$; for *setting I*, we have obtained $R_{wp} = 8.74\%$, $gof = 2.465$. The intensity mismatch found in a few peaks may be due to some poor grain statistics (reduced number of powder particles and significant size distribution) and to differences in site occupancies. Since these contributions are convoluted (for $0 < x < 5$) and we consider they are small, we chose not to refine them.

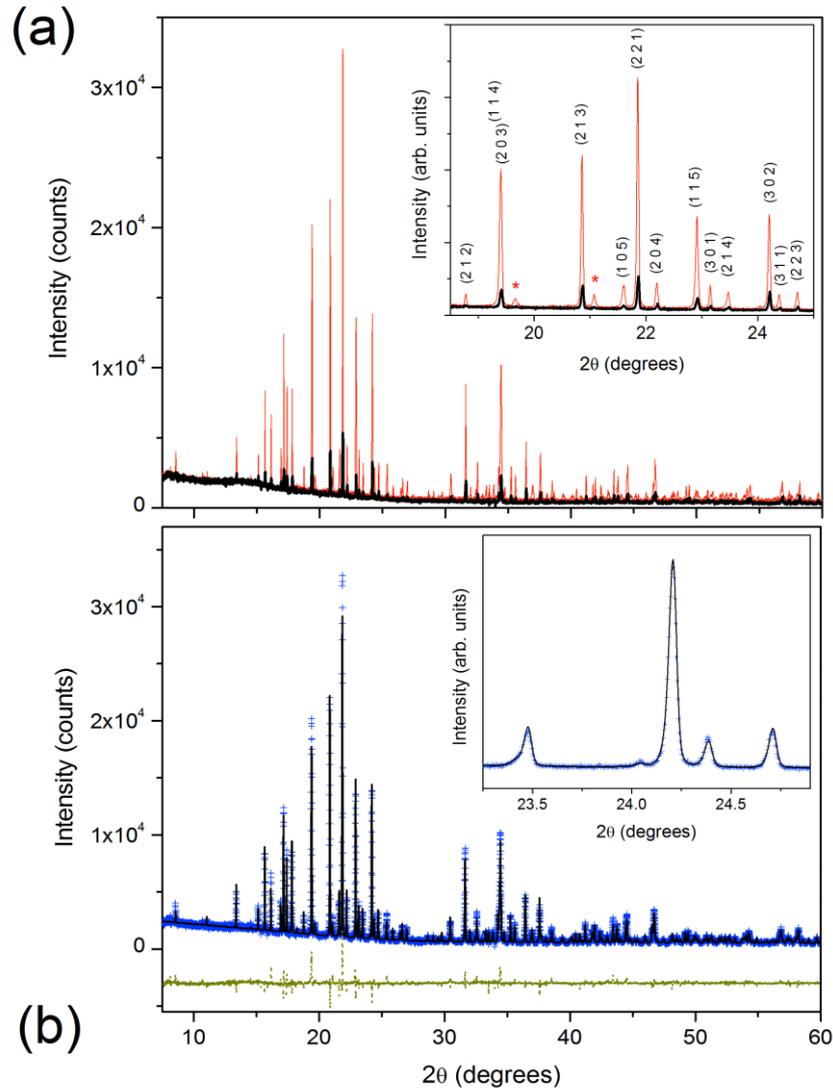

**Figure 1** (a) X-ray diffraction patterns for the powder of GdNd$_4$Si$_4$ compound fixed outside (thin line) and inside (thick line) the 0.3-mm-diameter capillary. Acquisition time: 360 s. The inset shows a few reflections of the tetragonal phase. The asterisks mark reflections of the secondary phase. (b) Observed and calculated X-ray diffraction intensities for the powder fixed outside the 0.3-mm-diameter capillary.

In Fig. 1a, we show the X-ray diffraction patterns for the powder of GdNd$_4$Si$_4$ compound fixed outside (*setting II*) and inside (*setting I*) the 0.3-mm-diameter capillary. Comparing the peak intensities, we observe that the signal/background ratio related to the maximum of the (2 2 1) reflection of the majority tetragonal phase (space group $P4_12_12$), for instance, is about 6.7 for the powder inside the 0.3-mm-diameter capillary and about 40.9 for the powder outside the same capillary, both measured with the same acquisition time of 360 s. In the inset of Fig. 1a, we can notice that the smallest peaks of the pattern of *setting II* are barely observed in the pattern of *setting I*. The differences between the patterns of *setting I* and *setting II* show the effects of the high X-ray absorption factor of this material at 12 keV (1.03 Å). For the compounds of the series $(Gd,Nd)_5Si_4$ measured inside a 0.3-mm-diameter capillary at 12 keV (1.03 Å), the absorption factor $\mu R > 14$.

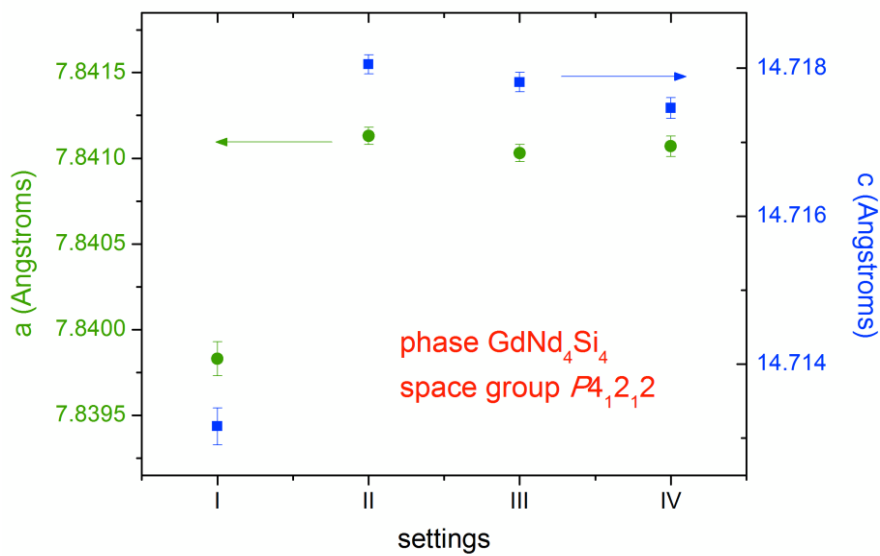

**Figure 2** Parameters *a* and *c* of the tetragonal phase in $GdNd_4Si_4$ compound obtained from different experimental settings: *setting I* - powder inside 0.3-mm-diameter capillary; *setting II* - powder outside 0.3-mm-diameter capillary; *setting III* - powder outside 0.5-mm-diameter capillary; *setting IV* - powder outside 0.7-mm-diameter capillary.

When we analyze the *e.s.d.* values from the refinements using *TOPAS* software, we observe that the *e.s.d.* values from pattern of *setting I* are always larger than the corresponding values from pattern of *setting II*, in spite of smaller *gof* value. This difference is shown in Fig. 2, where we present the parameters *a* and *c* obtained from Rietveld refinement for measurements in four different settings. It is easy to see that the *e.s.d.* values are larger for the refinement of the pattern of *setting I* in comparison with the other settings. Besides, considering the majority tetragonal phase (space group $P4_12_12$), the values of *a* and *c* parameters for *setting I* are different from those obtained for the other settings, and the values of *a* and *c* parameters for the settings *II*, *III* and *IV* are much closer comparing to each other. Then, considering the settings with the powder outside different capillaries, we observe a great reproducibility of the results, which are not dependent on the diameter of the capillaries, at least in the range between 0.3 and 0.7 mm. It is known that the capillary diameter influences the linewidths.

Indeed, taking the (2 3 1) reflection as an example, we observe that the linewidths are 0.041°, 0.044° and 0.049° for settings *II*, *III* and *IV*, respectively. The peaks present asymmetries that contribute to the refinement errors.

The methodology of measuring X-ray diffraction of the samples powder fixed outside the capillaries was applied to several representative compounds of the series $Gd_{5-x}Nd_xSi_4$ ($x$ = 1, 2, 3, 4 and 5). Analyzing the diffraction patterns, only $Gd_4NdSi_4$ did not have the secondary phase identified. The other compounds had the majority phase and the secondary phase adjusted, and the secondary phase is always the (Gd,Nd)Si orthorhombic phase (space group *Pnma*). The majority phase is the $(Gd,Nd)_5Si_4$ orthorhombic phase (space group *Pnma*) for the compounds with $x$ = 1 and 2. In the case of the compounds with $x$ = 3, 4 and 5, the majority phase is the $(Gd,Nd)_5Si_4$ tetragonal phase (space group $P4_12_12$). For $Nd_5Si_4$ compound ($x$ = 5), the lattice parameters of the tetragonal phase are $a$ = 7.8637(1) Å, $c$ = 14.8703(2) Å and $V$ = 919.55(3) Å$^3$. For the same compound and phase, Yang *et al.* (2002) reported the lattice parameters $a$ = 7.8694(0) Å, $c$ = 14.8077(1) Å and $V$ = 917.00 Å$^3$, while Roger *et al.* (2006) reported the lattice parameters $a$ = 7.8730(1) Å, $c$ = 14.8389(2) Å and $V$ = 919.78 Å$^3$. The relative amount of the phases are listed in Table I. The respective errors are the *e.s.d.* values from the refinements using *TOPAS*. The lattice parameters of the compounds are listed in Table II.

**Table I** Structural phase amounts and magnetic transition temperatures of compounds of the series $(Gd,Nd)_5Si_4$.

|  | $(Gd,Nd)_5Si_4$ tetragonal phase space group $P4_12_12$ % | $(Gd,Nd)_5Si_4$ orthorhombic phase space group *Pnma* % | (Gd,Nd)Si orthorhombic phase space group *Pnma* % | Magnetic transition temperature (K) |
|---|---|---|---|---|
| $Nd_5Si_4$ | 91.9(1) | - | 8.1(1) | 71 |
| $GdNd_4Si_4$ | 96.1(1) | - | 3.9(1) | 108 |
| $Gd_2Nd_3Si_4$ | 72.9(2) | - | 27.1(2) | 146 |
| $Gd_3Nd_2Si_4$ | - | 93.3(2) | 6.7(2) | 255 |
| $Gd_4NdSi_4$ | - | 100 | - | 297 |

**Table II** Structural parameters of $Gd_{5-x}Nd_xSi_4$ compounds obtained from Rietveld refinements.

| $x$ | 5 | 4 | 3 | 2 | 1 |
|---|---|---|---|---|---|
| Space group | $P4_12_12$ (#92) | $P4_12_12$ (#92) | $P4_12_12$ (#92) | $Pnma$ (#62) | $Pnma$ (#62) |
| a (Å) | 7.8637(1) | 7.8411(1) | 7.8121(1) | 7.5488(1) | 7.5154(1) |
| b (Å) | 7.8637(1) | 7.8411(1) | 7.8121(1) | 14.8665(2) | 14.8044(2) |
| c (Å) | 14.8703(2) | 14.7175(1) | 14.5969(2) | 7.8163(1) | 7.7864(1) |
| **Nd1, Gd1** | | | | | |
| Wyckoff | 4a | 4a | 4a | 8d | 8d |
| $x$ | 0.1901 | 0.1901 | 0.1901 | 0.36123 | 0.36123 |
| $y$ | 0.1901 | 0.1901 | 0.1901 | 0.25 | 0.25 |
| $z$ | 0 | 0 | 0 | 0.01188 | 0.01188 |
| $B_{iso}$ | 0.62(2) Å$^2$ | 1.20(2) Å$^2$ | 0.76(3) Å$^2$ | 0.41(3) Å$^2$ | 0.56(3) Å$^2$ |
| **Nd2, Gd2** | | | | | |
| Wyckoff | 8b | 8b | 8b | 8d | 8d |
| $x$ | 0.0142 | 0.0142 | 0.0142 | 0.03573 | 0.03573 |
| $y$ | 0.1326 | 0.1326 | 0.1326 | 0.09842 | 0.09842 |
| $z$ | 0.6192 | 0.6192 | 0.6192 | 0.18288 | 0.18288 |
| $B_{iso}$ | 0.62(2) Å$^2$ | 1.20(2) Å$^2$ | 0.76(3) Å$^2$ | 0.41(3) Å$^2$ | 0.56(3) Å$^2$ |
| **Nd3, Gd3** | | | | | |
| Wyckoff | 8b | 8b | 8b | 4c | 4c |
| $x$ | 0.9861 | 0.9861 | 0.9861 | 0.31107 | 0.31107 |
| $y$ | 0.3667 | 0.3667 | 0.3667 | 0.87892 | 0.87892 |
| $z$ | 0.2025 | 0.2025 | 0.2025 | 0.18149 | 0.18149 |
| $B_{iso}$ | 0.62(2) Å$^2$ | 1.20(2) Å$^2$ | 0.76(3) Å$^2$ | 0.41(3) Å$^2$ | 0.56(3) Å$^2$ |
| **Si1** | | | | | |
| Wyckoff | 8b | 8b | 8b | 8d | 8d |
| $x$ | 0.0529 | 0.0529 | 0.0529 | 0.2507 | 0.2507 |
| $y$ | 0.2810 | 0.2810 | 0.2810 | 0.25 | 0.25 |
| $z$ | 0.8070 | 0.8070 | 0.8070 | 0.3778 | 0.3778 |
| $B_{iso}$ | 0.62(2) Å$^2$ | 1.20(2) Å$^2$ | 0.76(3) Å$^2$ | 0.41(3) Å$^2$ | 0.56(3) Å$^2$ |
| **Si2** | | | | | |
| Wyckoff | 8b | 8b | 8b | 4c | 4c |
| $x$ | 0.2826 | 0.2826 | 0.2826 | 0.9831 | 0.9831 |
| $y$ | 0.3270 | 0.3270 | 0.3270 | 0.25 | 0.25 |
| $z$ | 0.6926 | 0.6926 | 0.6926 | 0.8992 | 0.8992 |
| $B_{iso}$ | 0.62(2) Å$^2$ | 1.20(2) Å$^2$ | 0.76(3) Å$^2$ | 0.41(3) Å$^2$ | 0.56(3) Å$^2$ |
| **Si3** | | | | | |
| Wyckoff | - | - | - | 4c | 4c |
| $x$ | - | - | - | 0.1365 | 0.1365 |
| $y$ | - | - | - | 0.9592 | 0.9592 |
| $z$ | - | - | - | 0.4756 | 0.4756 |
| $B_{iso}$ | - | - | - | 0.41(3) Å$^2$ | 0.56(3) Å$^2$ |

Even existing a change in the symmetry in the middle of the composition range of the series $Gd_{5-x}Nd_xSi_4$, the unit cell volume of the majority phase of each compound changes almost linearly as a function of neodymium content (Fig. 3a). A similar behavior is observed for the secondary phase. For this quasi-linearity, we assume that the majority phases (and the secondary phase, as a consequence) have the same Nd content of the nominal composition and it does not seem to depend on the secondary phase amounts. When we analyze the magnetic transition temperature ($T_C$), it is easy to see that there is a discontinuity in its variation as a function of Nd content (Fig. 3), just between the compositions where the majority phase changes from tetragonal ($x = 3$) to orthorhombic ($x = 4$). It is worth noting that the rate $dT_C/dx$ is almost the same for the tetragonal and orthorhombic phases: -19 and -20 K/ 10% *at*. Nd, respectively. The ratio of the majority orthorhombic phase was obtained including the data for $Gd_5Si_4$ compound from Pecharsky and Gschneidner Jr. (1997).

**Table III** The four smallest interatomic distances between different rare earth sites of the majority phase of each compound. These values were obtained using the software *Diamond* (version 3.2k).

|  | R1-R2a (Å) | R1-R2b (Å) | R1-R3a (Å) | R1-R3b (Å) |
| --- | --- | --- | --- | --- |
| $Nd_5Si_4$ | 3.488 | 3.496 | 3.667 | 3.684 |
| $GdNd_4Si_4$ | 3.471 | 3.478 | 3.655 | 3.656 |
| $Gd_2Nd_3Si_4$ | 3.454 | 3.460 | 3.631 | 3.642 |
| $Gd_3Nd_2Si_4$ | 3.536 | 3.592 | 3.469 | 3.475 |
| $Gd_4NdSi_4$ | 3.522 | 3.577 | 3.455 | 3.461 |

Considering the average of the four smallest interatomic distances (see Table III) between different rare earth sites of the majority phase of each compound, hereafter called *$4R_i$-$R_j$*, we can also observe a discontinuity in its variation as a function of Nd content (*x*) in the same *x* range observed for the magnetic transitions. To a better comparison, in Fig 3b, it is plotted the inverse of *$4R_i$-$R_j$* as a function of *x*, as well as the magnetic transition temperatures. It is easy to see the similarities between *($4R_i$-$R_j$)*$^{-1}$ vs. *x* and $T_C$ vs. *x*. Nevertheless, considering only two smallest distances between different rare earth sites, *($4R_i$-$R_j$)*$^{-1}$ vs. *x* behavior (not shown) is significantly different from the $T_C$ vs. *x* profile. In fact, a linear behavior is observed in the majority tetragonal phase, considering the nearest and next-nearest rare earth neighbors, which have the same distances, approximately (see Table III). Nevertheless, the linear behavior is only observed in the majority orthorhombic phase, when considering at least the three smallest interatomic distances between different rare earth sites. This is an indication that the exchange interactions (directly related to the magnetic transitions) in the orthorhombic phase are not governed by the distances among only the nearest and next-nearest rare earth neighbors of other rare earth ions in compounds of the series $Gd_{5-x}Nd_xSi_4$. The influence of further neighbors seems to be also very important. A similar assumption was done, for instance, in relation to compounds of the series $Nd_5Si_{4-y}Ge_y$ (Yang *et al.*, 2003). It is important to notice that the difference between the average distance of the third and fourth neighbors and the average distance of the first and second neighbors is smaller in the orthorhombic phase (~ 0.09 Å), by a factor of 2, comparing to the tetragonal phase (~ 0.18 Å).

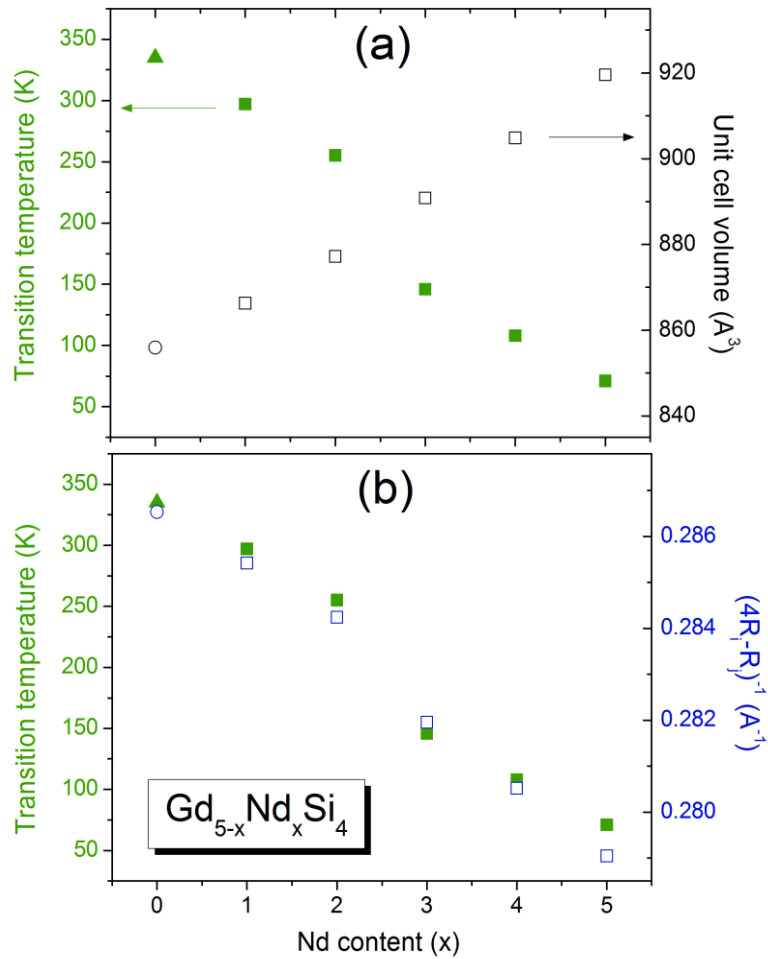

**Figure 3** (a) Magnetic transition temperature (solid symbols) and unit cell volume (open symbols) of the majority phase of each compound as a function of Nd content. (b) The inverse of the average of the four smallest distances between two rare earth sites in the majority phases (open symbols) as a function of Nd content in comparison with the magnetic transition temperatures (solid symbols). The error bars are smaller than the symbols. Solid triangles and open circles are data from Pecharsky and Gschneidner Jr. (1997).

## IV. Conclusion

In summary, we have demonstrated the efficiency of the XRD1 beamline to perform satisfactory X-ray powder diffraction on high-absorption materials, even not in optimal conditions. Applying the methodology of measuring the samples powder of compounds of the series $(Gd,Nd)_5Si_4$ outside the capillaries in Debye-Scherrer geometry at the XRD1, we have shown the results are satisfactory and reproducible using capillaries with different diameters (at least from 0.3 to 0.7 mm). Our analyses show that there is not a direct and simple correlation between the unit cell volume of the majority phases of the compounds and their magnetic transition temperatures. The results indicate that the behavior of the magnetic transition temperature as a function of Nd content may be directly related to the average of the four smallest interatomic distances between different rare earth sites of the majority phase of each compound, especially in the orthorhombic phase.

## Acknowledgements

The authors thank Dr. Cristiane B. Rodella, Henrique Canova, Adalberto Fontoura and Douglas Araújo for the previous and present efforts at the XRD1 beamline and Ana T. G. Mendes for arc-melting the samples. The authors also thank FAPESP (project number 2012/03480-0), CNPq and LNLS.